# Rigorous Description Of Design Components Functionality: An Approach Based Contract

Abdelhafid Zitouni[1]

[1] *Laboratory LIRE, Computer Science Department*
*Mentouri University of Constantine, Algeria*

**Abstract**
*Current models for software components have made component-based software engineering practical. However, these models are limited in the sense that their support for the characterization/specification of design components primarily deals with syntactic issues. To avoid mismatch and misuse of components, more comprehensive specification of software components is required,*
*In this paper, we present a contract-based approach to analyze and model the both aspects (functional and non-functional) properties of design components and their composition in order to detect and correct composition errors. This approach permits to characterize the structural, interface and behavioural aspects of design component.*
*To enable this we present a pattern contract language that captures the structural and behavioral requirements associated with a range of patterns, as well as the system properties that are guaranteed as a result. In addition, we propose the use of the LOTOS language as an ADL for formalizing these aspects. We illustrate the approach by applying it to a standard design pattern.*

## 1. Introduction

Component-based approaches have been proposed to create and deploy software systems assembled from components. The use of previously developed components should lead to faster time for complex software applications. Therefore, component-based software development is a promising solution to some of the problems that designers, developers and integrators face when building their systems [6]. Software patterns are a design paradigm used to solve problems that arise when developing software within a particular context. Patterns capture the static and dynamic structure and the collaboration among the different components in a software design. Since a design pattern is a recurring piece of software design, it can be seen as a component, called a design component in [15], and used to reify good design practice from conceptual design building blocks into a composable form. Design components focus on component-based problem solving instead of component-based implementation.
rigorous description of component functionality
The benefits of design patterns are that they serve as guidance to the novice designer, and they provide an extended vocabulary for documenting software design. Unfortunately, the descriptive format popularized by these catalogs is inherently imprecise. As a consequence, it is unclear when a pattern has been applied correctly, or what can be concluded about a system implemented using a particular pattern.

In order to address the ambiguity issues associated with design pattern descriptions, we introduce the concept of a design pattern contract as a formalism for precisely specifying design patterns. The responsibility of a pattern contract precisely characterizes the requirements that must be satisfied by the designer when applying a particular pattern.

Formal specification and verification techniques are useful for design analysis in that the sense are more precise, expressive, and unambiguous than the informal ones, such as graphical and textual notations. We argue that in order to achieve effective reuse it is important to specify both functional and architectural properties of a component in terms of formal specifications. Formal specifications are amenable to automation in analyzing component properties and thus facilitate the determination of reuse.

The formal description technique LOTOS (Language of Temporal Ordering Specifications) [4] was originally designed to specify the interactions among communicating processes, thus making it suitable for capturing the architectural (interaction) properties of components.

A contribution of this paper, is to provide a rigorous description of component functionality. This description can be achieved by means of contracts [18], using pre-and post-conditions for describing the semantics of component's services. Another contribution of this paper is a proposition of a novel Architecture Description Language (LOTOS-ADL) that has been designed to address specification of structural and dynamic architectures.

The rest of this paper is organised as follows. Section 2 introduces design patterns, and presents the concepts and notation of the LOTOS and contract. We present a short overview of our approach in section 3, before the main section –sect4- of this paper, we focus on the abstract specification of a component. Section 5 presents the concepts of LOTOS-ADL. Section 6 illustrate a case study and gives an overview of our environment of validation. Section 7 discuses the related work. Finally the last section concludes the paper and gives directions for future work.

## 2. Background

In this section, we introduce some basic concepts and terminology about design patterns, components, LOTOS and design by contract.

2.1 Design Patterns

Design patterns are a design paradigm used to solve problems that arise when developing software within a particular context [10]. Patterns capture the static and dynamic structure and collaboration among the components in a software design. To build software systems, a designer needs to solve many problems. Applying known design patterns to address these problems allows the designer to take advantage of expert design experience documented in each pattern. Although design patterns are not formal in nature, design components that have been inspired by design patterns are amenable for formal modeling and analysis. The focus on design components is important because one of the goals of our work is to detect errors as early as possible in the development process by reasoning about the properties at the design level and reducing the cost of finding and correcting these errors in concrete software components.

2.2 LOTOS

LOTOS is a formal description technique based on a combination of CCS [19] and CSP [14]. In LOTOS, a system is seen as a process, possibly consisting of several sub-processes. Likewise a sub-process is a process in itself, and a LOTOS specification describes a system via a hierarchy of process definitions. A process is an entity capable of performing internal, unobservable actions, and of interacting with other processes which form its environment. In that sense, LOTOS implements a black box paradigm used to develop high level, concise and abstract specifications of complex systems. At some abstraction level, it is possible to express the interactions of a process with its environment without having to describe the internal structure (or implementation) of that process. Process definitions are expressed by the specification of behaviour expressions that are constructed by means of a restricted set of powerful operators making it possible to express behaviours as complex as desired.

Basic LOTOS is a subset of LOTOS. The processes interact with each other by pure synchronization without exchanging any value. Fig.1 provides an intuitive illustration of the main Basic LOTOS operators.

| Operator | Description | Example |
|---|---|---|
| [ ] | Either P1[a,b] or P2[c,d] depending on the environment | P[a,b,c,d]=P1[a,b] [ ] P2[c,d] |
| ||| | Parallel composition without synchronization: P1[a,b] is independent from P2[c,d] | P[a,b,c,d]=P1[a,b] ||| P2[c,d] |
| [b] | Parallel composition with synchronization on gate b | P[a,b,c,]=P1[a,b] \|[b]\| P2[b,c] |
| | Parallel composition with synchronization on several gates (b,c,d) | P[a,b,c,d,e]=P1[a,b,c,d] \|[b,c,d]\| P2[b,c,d,e] |
| hide b in \|[b]\| | Parallel composition with synchronization on gate b, moreover where gate b is hidden | P[a,c]=hide b in P1[a,b]\| [b]\| P2[b,c] |
| >> | Sequential composition P1[a,b] is followed, when P1 termited, by P2[c,d] | P[a,b,c,d]=P1[a,b] >> P2[c,d] |
| [> | Disrupt: P1 [a, b] may be interrupted at any time before its termination by P2[c, d]. | P[a,b,c,d]=P1[a,b] [> P2[c,d] |
| ; | Process prefixing by action a | a;P |
| Stop | Process which cannot communicate with any other process | Stop |
| Exit | Process which can terminate and then transforms itself into stop | Exit |

Fig.2. Basic LOTOS operators [3]

2.3 Design by contract

Design by contract is a design approach developed by Meyer [18]. It is used here to provide precise specifications for the functionality of components and to enhance their reliability. According to Meyer, a contract is a collection of assertions that describe precisely what each feature of the component does and does not do. The key assertions in the design by contract technique are of three types: invariants, pre-conditions, and post-conditions. An invariant is a constraint attached to type that must be held true for all instances of the type whenever an operation is not being performed on the instance.

Pre-conditions and post-conditions are assertions attached to an operation of a type. A pre-condition expresses requirements that any call of the operation must satisfy if it is to be correct. A post-condition expresses properties that are ensured in return by the execution of the call.

## 3. Overview of the Approach

In [27] we have presented a systematic approach for a software designer to model and analyze component integration during the design phase, the early planning stage of the software lifecycle. This approach includes a process of representing, specifying, instantiating and integrating design components and analyzing their compositions, which are captured as contracts. The process is illustrated in Fig.2.

This approach allows design components to be reused by making the components description available in a component library. With this approach, the designer can not only model the design component precisely, unambiguously and expressively, but also detect the interactions between components and correct design errors before implementation [26]. As shown in figure 2, our approach begins by four steps: (Analysis, selection, abstract specification and the instantiation steps).

In this article we focus on the abstract specification of the component and the ADL for describing architecture of component-based software, which provide explicit support for specifying components. ADLs are important since they can document component-based architecture early, reason about their properties, and automate their analysis and system generation [12].

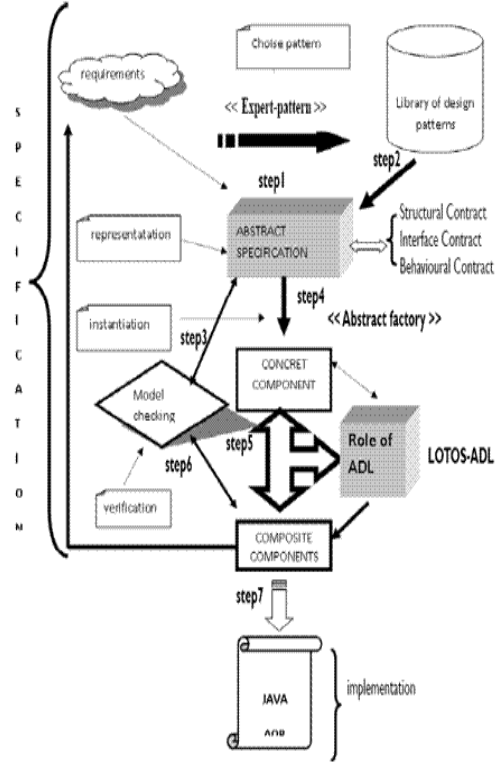

Fig.2. Overview of our approach [27]

## 4. Abstract specification of a component

The abstract specification is inspired from the work of Dong and al. [7]) and contains a formal model of design component, called design component contract. A design component contract includes structural contract (SC), behavioural contract (BC) and interface contract (IC).

The structural properties describe the relations of the constructs of each design component. The behavioural properties are constraints such as event ordering, and action sequence of each design component. The interface contract describes the finite set of input or output ports attached to a design component and the set of messages sent to or received by a component. We define an abstract specification contract (ASC) as::

**ASC**::=<Component-Name>**Where**<assertion>**and**

<SC>**and**<IC>**and** <BC> **End**

## 4.1 A motivating Example

To motivate this paper we consider the structure (class and interaction diagram) of the Observer pattern shown in fig. 3 [10]: (The Observer pattern (also called Publisher-Subscriber) regulates how a change in one object can be reflected in an unspecified number of dependant objects).

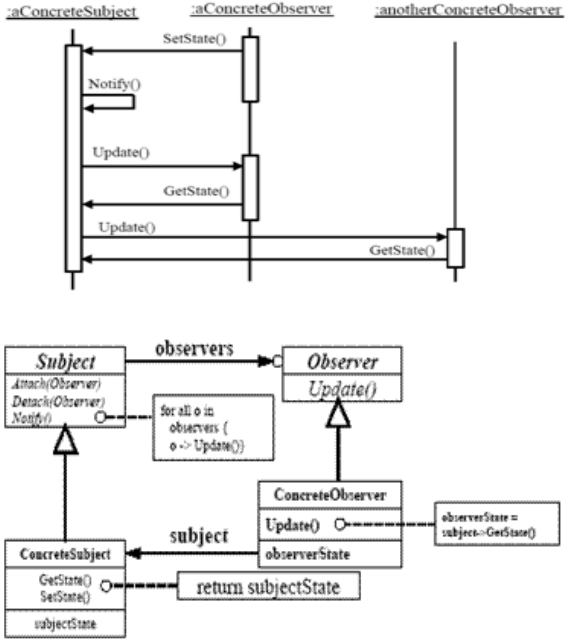

Fig. 3  Observer pattern (class diagram, interaction diagram)

## 4.2 Structural contracts

In [27] we have formalized the structural aspect of a design component contract by using a subset of First Order Logic (FOL), because the relations between pattern participants can be easily expressed as predicates. The subset of FOL used to describe the structural aspect of a design component comprises variable symbols, connectives ('∧'), quantifiers ('∃'), element (ϵ) and predicate symbols acting upon variable symbols. The variable symbols represent class, objects, while the predicate symbols represent permanent relation [24].
We define two groups of predicates, entities (Table 1) and relationships (Table 2).
- Entity predicates define whether a design component has a specific class (abstract or concrete), what a method (or attribute) is defined in a class….
- Relationship predicates define the relations between classes, attributes, and operations and the actions that a role can perform in a component.

The Abstract specification of a component presented in this paper is an extension of the model existing domain in [27] by introducing the concepts related to the Aspect-Oriented Approach.
For the concepts related to the Aspect-Oriented approach, we will define new predicates. (Table 1) (Table 2)
An Aspect-Class Diagram contains the classes, the aspects and the interfaces, linked with relations, which include associations, generalization, and realization between classifiers and calls between operations.

Table1: Entity predicates

| Predicate | Description |
| --- | --- |
| Abstract-class ( C ) | C plays the role as an abstract-class in the component |
| Abstract-Aspect(A) | C plays the role as an abstract-aspect in the component |
| Class ( C ) | C plays the role as concrete class in the component |
| Aspect(A) | A plays the role as concrete aspect in the component |
| x ϵ X | X is an element of set X |

Table2: Relationship predicates

| Predicate | Description |
| --- | --- |
| Inherit (A,B) | B is a subclass of A |
| Associate (A,B) | A,B are connected with association relation |
| Aggregate (A,B) | A contain a reference to B |
| Invoke (A,m1,B,m2) | A method m1 defined in class A calls a method m2 defined in class B |
| New(A,m,O) | The method m of class A create a new object of type A |
| Return (A,m,O) | The method m of class A returns an object O of type A |
| DeclareParent (A,B,C) | B is a sub class of A. This relation is declared in aspect C |
| call (A,cp,B) | The Pointcut CP of aspect A designates a set of join points of the class B |
| advice(A,cp,action) | Advice codes implement the behavior of an aspect A. Several types of action exist: before, around, after returning,. |

## 4.3 Interface contracts

We define the interface aspect of a design component contract as follow:

Let a tuple IC = (P, IP,OP, IM,OM, IMI ), where P is a finite set of process names, IP is a finite set of input ports attached to a process, OP is a finite set of output ports attached to a process, IM is a finite set of input messages sent to a process and OM is a finite set of output messages sent from a process, IMI is the finite set of input messages sent from outside the design component to a process.

The abstract specification of the interface contract of Observer is done by:

(0) Component-name is Observer where:
(1)    $\exists$ (aConcreteSubject, aConcreteObserver, anotherConcreteObserver) $\in$ C
(2)  $\wedge$  $\exists$ ( inOS, inSO,self, input) $\in$IP
(3)  $\wedge$  $\exists$ (outOS, outSO, output ) $\in$OP
(4)   $\wedge$    $\exists$ (attach, detach, getstate, setstate,update, notify, change ) $\in$ IM
(5)  $\wedge$  $\exists$ (attach, detach, getstate, setstate, update, notify) $\in$ OM
(6)  $\wedge$   $\exists$ (change) $\in$ IMI

In order to be able to support dynamic reconfiguration of the service and to provide precise specification about the relationships of operations calls to each other, we include the constraints on component interfaces.

This allows assertions about the gates (set of input or output ports attached to a process) to appear in pre-conditions, and post-conditions.

Let IC1 = (IC, Constraint) we denote:
$p \in$ IP(p) = {i $\in$ IP \ gate_Ini = p} $\wedge$
$p \in$ OP(p) = {i $\in$ OP \ gate_Outi = p} $\wedge$
$m \in$ IM(p) = {i $\in$ IP, m $\in$ IM \ gate_Ini ?m} $\wedge$
$m \in$ OM(p) = {i $\in$ OP, m $\in$ OM \ gate_Outi !m} $\wedge$
a ll-gateIN= { all IP(p)/ p $\in$ Component } $\wedge$
all-gateout={ all OP(p)/ p$\in$ Component } $\wedge$
**Where** Constraint /*constraints on gates*/:

$\forall i,j \in 1,n \rightarrow$ gate_Ini $\neq$gate_Inj $\wedge$
$\forall i,j \in 1,n \rightarrow$   gate_Outi  $\neq$  gate_Outj   $\wedge$
$\forall j \in 1,n \rightarrow \exists i \in 1,n$/gate_Inj ?mi $\in$ gate_Outi !m i $\wedge$
$\forall j \in 1,n \rightarrow \exists i \in 1,n$/ gate_Outj?mi $\in$ gate_Ini !m

.

## 4.4 Behavioural contracts

In contrast to the structural aspect of a design component contract, the behavioural contract describes the dynamic information, such as the collaboration among the objects participating in the component and the creation of new objects.

We have chosen a basic LOTOS for defining a formal semantic model of behavioural contracts because it represents a powerful approach for modeling behaviour and concurrency. The choice of LOTOS is motivated by its powerful ability for describing behaviour and the availability of tools enabling formal verification and automatic generation of distributed programs. Our proposal focuses on formally describing architectures encompassing both the structural and behavioural viewpoints. The LOTOS specification of the observer follows:

**Specification** Observer [input,output] **: noexit:=**
    /*…. Signature……*/
**behaviour**
   aConcreteSubject [input, output]
        |[input, output]|
   aConcreteObserver [input, output]
            [ ]
   anotherConcreteObserver [input, output]
  **where**
 **Process**  aConcreteSubject [inCS, outCS]:= **noexit**
        ?setstate; !notify; !update ;?getsate;
        aConcreteSubject [inCS, outCS]
**Endprocess**
**Process** aConcreteObserver [inaCO, outaCO] **:= noexit**
       I; !setstate; ?update; !getstate
        aConcreteObserver [inaCO, outaCO]
**Endprocess**
**Process**   anotherConcreteObserver[inbCO,outbCO]   **:= noexit**
       I; !setstate; ?update; !getstate
        anotherConcreteObserver [inbCO, outbCO]
**Endprocess**

**Endspec**

## 5. Proposal Architecture Description Language

A key aspect of the design of any software system is its architecture. From a runtime perspective, an architecture description should provide a formal specification of the architecture in terms of components and connectors and how they are composed together. Enabling specification of dynamic architectures is a large challenge for an Architecture Description Language (ADL). This section describes LOTOS-ADL, our proposal ADL that has been designed to address specification of structural and dynamic architectures. While most ADLs focus on defining software architectures from a structural viewpoint, our proposal LOTOS-ADL focuses on formally describing architectures encompassing both the structural and behavioural viewpoints.

From a runtime perspective, two viewpoints are frequently used in software architecture: the structural viewpoint and the behavioural one.

The structural viewpoint may be specified in terms of: components, connectors, and configurations of components and connectors.

   **<LOTOS-ADL>**:= < structural viewpoint, behavioural viewpoint>;
   < structural viewpoint> := <component, connector, configuration>/
     **component** := <cp1, cp2, ....., cpn>  n ≥ 2 and
     **connector** := <$ct_1$, $ct_2$, ....., $ct_m$>   m ≥ 1
    with **constraints:**
 ∀cp1, cp2∈ component / name.cp1 ≠name.cp2
 ∀ct1, ct2∈ connector / name.ct1 ≠name.ct2
    **configuration:** = < /* LOTOS operators construct*/>
   <behavioural viewpoint>:=  < LOTOS behavior expression >

Thereby, from a structural viewpoint, an architecture description should provide a formal specification of the architecture in terms of components and connectors and how they are composed together. Further, in the case of a dynamic architecture, it must provide a specification of how its components and connectors can change at runtime. The behavioural viewpoint may be specified in terms of: actions a system executes or participates in, relations among actions to specify behaviours, and behaviours of components and connectors, and how they interact.

A LOTOS specification describes a system through a hierarchy of active components, or processes. A process is an entity able to realize non-observable internal actions, and also interact with others processes through externally observable actions.

 We model a component as a black-box with a set of input and output gates (or channels), where visible events occur.

Instead of describing the static functionalities that a component provides, we specify the set of (dynamic) behaviors that a component may exhibit in constituting a system. All gates, together with constraints that may be imposed upon the ports, constitute the interface of a component. The interface of a component specifies the constraints on the way the component is to be used. A component may have overall constraints imposed upon the gate. The set of concepts that are manipulated are presented within our ADL meta-model (Fig. 4).

In our meta-model, we are mainly interested in representing static and dynamic behaviour contract using static and dynamic contract. A major benefit of separate static part from the dynamic part is that reasoning independently from any particular situations. The static contract of a component is a part that does not evolve. The evolution of a dynamic contract may have different purposes.

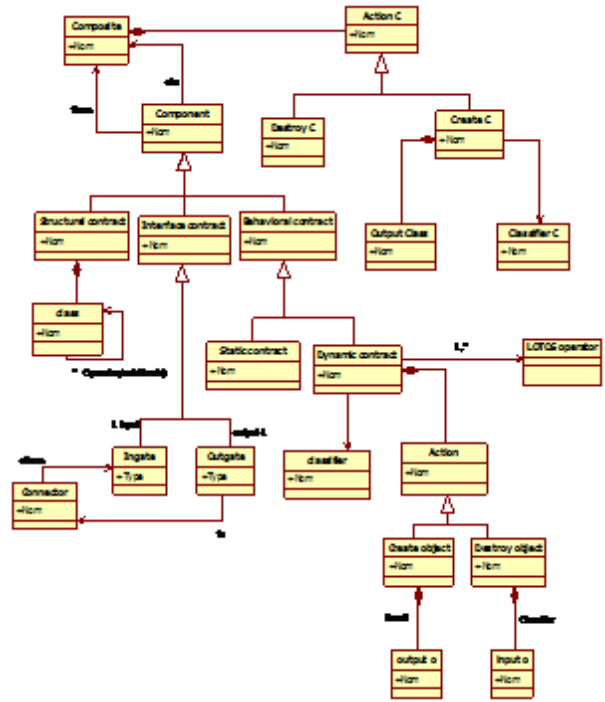

Fig.4. The LOTOS-ADL Meta-model

## 6. Case study: Client/server

Let us, consider the simple client-server system shown in Figure5. It consists of one client and one server interacting via link connector. Such a system is easy to describe in LOTOS-ADL. A LOTOS-ADL specification describes a system through a hierarchy of components (process). A process is an entity able to realise non-observable actions, and also interact with others process through externally observable actions.

The LOTOS specification at the top-level is a parallel composition of the process Client (component client), the process Server (component server) and the process connector (connector) (Fig.5). In order to specify this system, we adopt the following guiding [22]:
- The basic architecture elements, namely components and connectors, are modelled through the basic LOTOS abstraction, namely process.
- Any two LOTOS processes that model components must be in parallel composition with a LOTOS process defined as a connector
- The service specification consists of the temporal ordering of events executed at the service interface.
- We call to invocation (inv) those actions to activate the service and termination (ter) to the action of return a result.

### 6.1. Point to point connector

The LOTOS specification at the top-level is a parallel composition of the process Client (component client), the process Server (component server) and the process connector (connector) (Fig.5).

  **specification** Client-Server [invClt,terClt,invSrv,terSrv] **: noexit:=**
     **library** RESULT, SERVICES **endlib**
     **behaviour**
      Client [invClt, terClt]
         |[invClt, terClt]|
      connector [invClt, terClt, invSrv, terSrv]
         |[invSrv, terSrv]|
      Server [invSrv, terSrv]
     **where**
       ………
       ………
     **Endprocess**

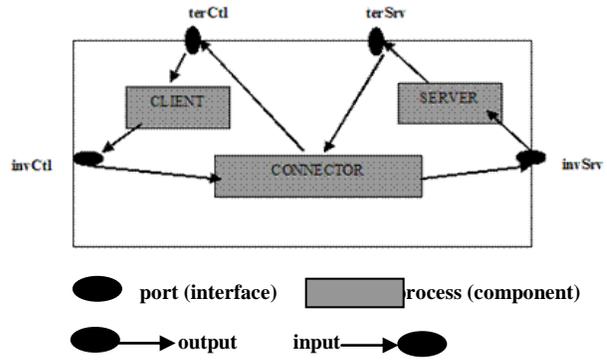

port (interface)     rocess (component)
output     input

Fig. 5. Illustration of the Client-Server specification

The connector behaviour is defined through the temporal ordering of invocation operations in the connector interface. The connector interface is made up of four ports: invCtl to invocations from client, terCtl to returns to client, invSrv to invocations from server and terSrv to return to server

**process** Connector[invClt,terClt,invSrv,terSrv] **: noexit: =**
    invClt ? s : SERVICE ? op: OPER     /* the client passes the request to connector*/
    invSrv ! s ! op; /  * the connector passes the request to the server*/
    terSrv ! s ? r : RESULT;   /*the server passes the reply to the connector*/
    terClt ! s ! r;   /*the connector passes the reply to the client*/
      Connector [invClt, terClt, invSrv,terSrv]
  **Endproc**

In this case, the connector receives an invocation from the server that contains both the name of the requested service and the operation being requested on the server (invClt?s: SERVICE? Op: OPER). The connector passes both of them to the server and waits for the reply. Finally, the connector passes the reply containing the result to the client.

### 6.1. Multicast connector

The connector abstract software architecture is defined as a collection of services. In order to specify the connector abstract software architecture, we assume that is composed by three components (Fig. 6) (service1, service2, service3) and a single connector (communication Service). The LOTOS specification of this software architecture is done by a parallel composition of the set of basic services and the process Communication Service.

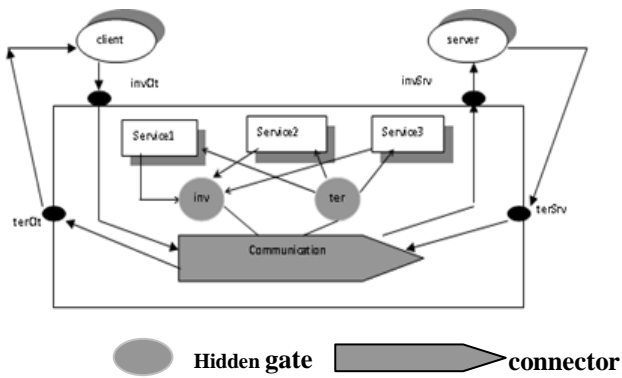

**Hidden gate**    **connector**

Fig. 6.  Illustration of the abstract software architecture

**Process**Connector_Abstract[invClt, terClt, invSrv, terSrv] **: noexit : =**
  **hide** inv, ter **in**
  ((Service1 [inv, ter ] |||
    Service2 [inv, ter ] ||| Service3 [inv, ter ])
                ||
  ServiceOrdering [inv, ter ])
        |[inv,ter]|
CommunicationService[inv,ter,invClt,terClt,invSrv,terSrv]
  **Where**
  ……
  ……
  **Endproc**

According to the constraints imposed by ServiceOrdering, after the request gets in the connector, it is passed to Service1 followed by Service2 and Service3. The LOTOS specification of the ServiceOrdering is done by:

  **Process** ServiceOrdering [inv,ter] **: noexit : =**
  inv ! Service1 ? op: OPER
  ter ! Service1 ? r : RESULT
  inv ! Service2 ? op: OPER
  ter ! Service2 ? r : RESULT
  inv ! Service3 ? op: OPER
  ter ! Service3 ? r : RESULT
    ServiceOrdering [invClt, terClt, invSrv,terSrv]

  **Endproc**

## 6. Verification

For the verification of our approach, we use our environment of verification, named FOCOVE (Formal Concurrency Verification Environment) [27] (available in www.focove.new.fr)    (Fig. 7). Focove is an integrated environment designed to edit Basic LOTOS  behavior expressions which describe reactive systems and to generate and analyze Maximality based Labelled Transitions Systems structures (MLTS).   concerns the state of the art of ADLs.

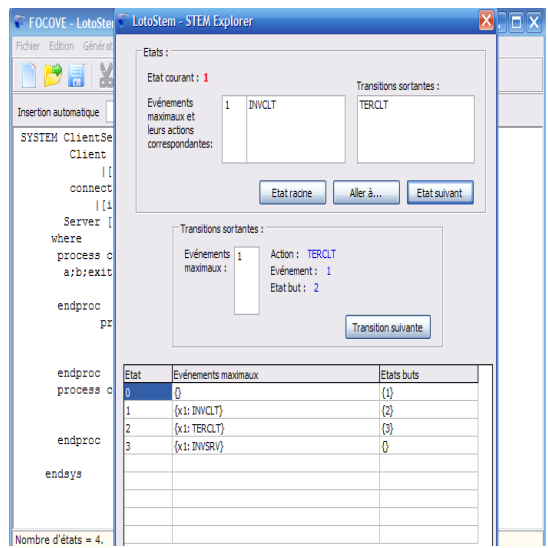

Fig 7. The environment of verification

The FOCOVE environment is dedicated to the design and verification for component based software development. FOCOVE translates a LOTOS program into a Labelled Transition System (LTS for short) describing its exhaustive behaviour. This LTS can be represented either explicitly as a set of states and transitions or implicitly as a library of C functions allowing us to execute the program behaviour in a controlled way.  By verification, we mean comparison of a complex system against a set of properties characterizing the intended functioning of the system (for instance, deadlock freedom, mutual exclusion, etc.).

## 7. Related works

We have chosen three dimensions to compare our approach with other existing work. The first dimension concerns the use the design pattern in designing applications. The second dimension concern of application of the contract notion and the approach to software component specification. The third dimension concerns the state of the art of ADLs.

-Several successful experiences have reported on the advantages of using patterns in designing applications [11], [23]. These experiences do not follow a systematic method to develop applications using patterns. Systematic development using patterns utilizes a composition mechanism to glue patterns together at the design level.

Generally, we categorize composition mechanisms as behavioural and structural compositions. Behavioural composition approaches are concerned with objects as elements that play several roles in various patterns. Reenskaug [20] developed the Object Oriented Role Analysis and Software Synthesis method. The method uses a role model that abstracts the traditional object model. Riehle [21] uses role diagrams for pattern composition. The approach by Jan Bosch [4] uses design patterns and frameworks as architectural fragments. Each fragment is composed of roles and components that are merged with other roles to produce application designs.

-The notion of contracts in software development is attributed to Meyer [18]. Another contribution, the object-oriented contracts of Helm et al. [13] focused on specifying the behaviour and interactions between objects in a system. Helm et al. noticed that the behaviour of an object could not be inferred from its interface, leading to design and reuse problems. Contracts formalize the behavioural relationship between objects and define a set of participants and their obligations. In this paper, we defined a formal model of design component based on contracts and a rigorous analysis approach to software design composition.

Keller and Schauer [15] described a methodical approach to design composition which was illustrated as a process within a four-dimensional design space. They characterized a special kind of component, called a design component, and discussed a development process to compose these components at the design level and generate source-code frames or executable code. Although our approach is also in the area of software composition, it focuses on the formal, declarative, and property-based aspects of design composition.

- The majority of the ADLs support only a structural view of the system. Even if offering any techniques for describing behaviour of the system, they only model its possible behaviour and thus can check its consistency only statically (e.g. correctness of proposed configuration, type checking, pre- or post-conditions, protocol). A few of them support dynamic configurations. C2 [25] specifies only pre- and post-conditions, Darwin [17] expresses component semantics in terms of π-calculus. Weaves [12] defines partial ordering of data-flow over input and output objects, but only Rapide [16] and Wright [2] specify dynamic component behaviour. Wright focuses on specifying communication protocols among components and uses a variant of CSP [14] to describe architectural behaviour. It treats both components and connectors as processes, which synchronise over suitably renamed alphabets. But, it implies a component interface extension in case of permitted reconfiguration and checks only if a connector protocol is deadlock-free as a consistency check [1]. Moreover, none of these ADLs have component have first class in order to cope with description of dynamic and mobile architecture.

**4. Conclusions**

In this paper, we have introduced a proposition of formal model of design component based on contract and a rigorous analysis approach to software design composition based on automated verification techniques. Our approach allows us to find errors in the design composition early in the development process. This paper has illustrated how to adopt LOTOS as ADL to describe the behaviour of software architecture.

This language is mathematically well-defined and expressive: it allows the description of concurrency, non-determinism, synchronous and asynchronous communications. It supports various levels of abstraction and provides several specification styles. These positive features encouraged us to adopt LOTOS as an ADL for describing both component and connector enables us to check behaviours properties. Finally, LOTOS specifications can also be used to express and verify concurrency models and real-time properties of systems.

The presented LOTOS specifications serve as a basis for very interesting future work. We are currently interested in the refinement of specifications in which the refinement process follows the rules of the software architecture refinement.

Also, we are investigating to proposing a rules-based transformation enabling the mapping from LOTOS specification to JAVA pseudo code.

**References**
[1] R. Allen, D. Garlan, and R. Douence.: Specifying dynamism in software architectures. In Proceedings of the Workshop on Foundations of Component-Based Software Engineering, Zurich, Switzerland, September 1997.
[2] R. Allen.: A Formal Approach to Software Architecture. PhD thesis, Carnegie Mellon, School of Computer Science, January 1997.: Issued as CMU Technical Report CMU-CS-97-144.

[3] Aprille L, Saqui-sannes P, Lohr C.: A new UML profile for reel-time system formal design and validation,. in LNCS 2185, 2001

[4] T.Bolognesi, E.Brinksma. Introduction to the ISO specification language LOTOS . In Van EIJK, pp 23-73, 1989

[5] J. Bosch.: Specifying Frameworks and Design Patterns as Architecture Fragments. Proceedings of Technology of Object-Oriented Languages and Systems,China, Sept. 22-25 1998.

[6] Jing Dong.: Design component contracts, Phd thesis. Computer Science department, university of Waterloo, June 2002.

[7] Dong J, Paulo S C Alencar, Donald D Cowan.: Automating the analyse of design component contracts, In software Practice and Experience, 2005.

[8] Dong, J., Yang,S., Huynh, D:. Evolving Design Patterns Based on Mode Transformation, *Proceedings* of the Ninth IASTED I C S and Applications (SEA), pp 344-350, USA 2005.

[9] Ehrig,H., Mahr,B.,. Fundamentals of Algebraic specification, volume1, Springer-verlag, Berlin., 1985

[10]. Erich Gamma, Richard Helm, Ralph Johnson, John Vlissides.: Design Patterns, Elements of Reusable Object-Oriented Software, , Addison-Wesley Longman. 1995

[11] J. Garlow, C. Holmes, T. Mowbary.: Applying Design Patterns in UML. Rose Architect, Vol 1, No. 2, Winter 1999

[12] M.M. Gorlick, R. Razouk.: Using Weaves for software construction and analysis. In Proceedings of the 13th international conference on Software engineering, pages 23-34. IEEE Computer Society Press, 1991.

[13] R. Helm, I.M Holland, D. Gangopadhyay.: Contracts: Specifying behavioral compositions in object-oriented systems.Proceedings of the ACM Conference on Object-Oriented Programming Systems, Languages and Applications (OOPSLA),October 1999;

[14] T. Hoare.: Communicating Sequential Processes. Prentice Hall International, 1985.

[15] R. K. Keller, R. Schauer. Design Components: Towards Software Composition at the Design Level. *Proceedings of the 20th International Conference on Software Engineering*, pages 302–311, 1998.

[16] D.C. Luckham, J. Vera.: An event based architecture definition language. IEEE Transactions on Software Engineering, 21(9):717-734, September 1995.

[17] J. Magee, N. Dulay, S. Eisenbach, and J. Kramer.: Specifying distributed software architectures. In Proc. of 5th European Software Engineering Conference(ESEC'95),,pages 137-153. Springer-Verlag, September 1995.

[18] Meyer B.: Applying 'design by contract'. IEEE Computer:40-51, October 1992

[19] R. Milner,: Communication and Concurrency, Prentice Hall, Englewood Cliffs, NJ, 1989.

[20] T. Reenskaug.: OORASS: Seamless Support for the Creation and Maintenance of Object Oriented Systems. Journal of Object Oriented Programming, 5(6):27-41, October1992.

[21] D.Riehle.: Composite Design Patterns. Proceedings of Object-Oriented Programming, Systems, Languages and Applications,OOPSLA'97, pp218-228, Atlanta, October 1997.

[22] N S Rosa, Paulo R Cunha,: A software architecture-based approach for formalising middleware behavior in ENTCS 2004.

[23] S. Srinivasan , J. Vergo.: Object-Oriented Reuse: Experience in Developing a Framework for Speech Recognition Applications. Proceedings of 20[th] International Conference on Software Engineering,ICSE'98, pp322-330, Kyoto, Japan, April 19-25, 1998.

[24] Taibi T. and Ngo D.C.L .: Modeling of distributed objects computing design patterns combination. Journal AMCS vol 13 N° 2 pp 239-253, 2004

[25] R.N. Taylor, N. Medvidovic, K.M. Anderson, E.J. Whitehead Jr., J.E. Robbins, K.A. Nies, P. Oreizy, D.L. Dubrow.: A component and message based architectural style for GUI software. IEEE Transactions on Software Engineering 22(6):390-406, 1996.

[26] A. Zitouni. : Un framework pour l'utilisation des design patterns par intégration du langage de spécification LOTOS, Congré International en Informatique Appliquée CIIA05, novembre 2005, BBA, Algérie, ISBN: 9947-0-1042-2

[27] Zitouni, A., Seinturier, L., Boufaida.,M., 2008. Contract-based approach to analyze software components, International Conference on *Engineering of Complex Computer Systems* (ICECCS 2008/UML&AADL) workshop, Belfast April, pp 237, 242.

[28] A. Zitouni, M. Boufaïda, L. Seinturier, "Specifying Components With Compositional Patterns, LOTOS and design by contract". ISCA, 19[th] International Conference on Software Engineering and Data Engineering (SEDE-2010), , San Francisco, California, USA, pp 190-195. June 16 - 18, 2010

**Author** Dr. Abdelhafid Zitouni received his PhD in computer science in 2008 from the University Mentouri of Constantine, Algeria. Currently working as Assistant professor in Mentouri University of Constantine. His research interests include Software Engineering, Software Design, Software Reuse and Design pattern Detection, formal methods in software.